# Composition, structure, and luminescent properties of $SiO_xN_y(Si)$ composite layers containing Si nanocrystals

V.G. Baru[1], I. N. Dyuzhikov[1], V. I. Pokalyakin[1], O. F. Shevchenko[1], E. A. Skryleva[2], O. M. Zhigalina[3]

[1]Institute of Radioengineering and Electronics of RAS, Moscow, Russia
[2]Moscow Steel and Alloys Institute, Moscow, Russia
[3]A.V. Shubnikov Institute of Crystallography of RAS, Moscow, Russia

A relationship between the chemical composition, structure and luminescent properties of light-emitting $SiO_xN_y(Si)$ composite layers with Si nanocrystals is demonstrated. Photoluminescence (PL) with a maximum of intensity at $\lambda \approx$ 500-600 nm is observed in a narrow region of chemical compositions with relatively small Si excess ( $a \approx$ 10 at. %). Composite layers structure is studied by means of HRTEM. Appearance of nanocrystals due to annealing is accompanied by substantial growth (30-40 times) of PL intensity but do not change PL spectra shape. Chemical composition of structural luminescent-active complexes with excess Si atoms is determined by XPS technique.

## Introduction

The development of light-emitting silicon nanostructures, study of their properties as well as identification of physical mechanisms of photo- (PL) and electroluminescese (EL) are of great interest because of certain success in a way to Si laser and LED fabrication [1]. In this work we demonstrate a strong relationship between the chemical composition, structure and luminescent properties of $SiO_xN_y(Si)$ composite layers with Si nanocrystals.

$SiO_xN_y(Si)$ composite layers were obtained using ion-plasma sputtering of Si target in the atmosphere of $N_2$ and $O_2$ followed by high-temperature annealing. The study of elemental and phase composition of the layers as well as the study of chemical composition of luminescent complexes was performed by means of X-ray Photoelectron Spectroscopy (XPS) using spectrometer PHI 5500 ESCA. The structure of the layers have been investigated using high-resolution transmission electronic microscopy (HRTEM) with the help of microscope Philips EM 430 ST. PL spectra in the visible region are measured by an imaging monochromator SolarTII MS3504i with Si CCD detector. A laser with wave length $\lambda$= 355 nm was used for PL spectra excitation.

## Results

It was found that intensive PL arises in the narrow range of chemical compositions with the relatively small over-stoichiometric Si excess $a$ in investigated layers ( $a \approx$ 10 at. %). Fig. 1 shows PL spectra of four samples after annealing in vacuum ($T$ = 800 C, $t$ =30 min). Inset in Fig. 1 shows the dependences of PL intensity in spectra maximums for these samples on Si excess. PL spectrum shape is slightly dependent on $a$, while PL intensity demonstrates strong non-monotone dependence on $a$. Further increase of $a$ ( $a >$ 13 at. %) leads to strong decrease of PL intensity.

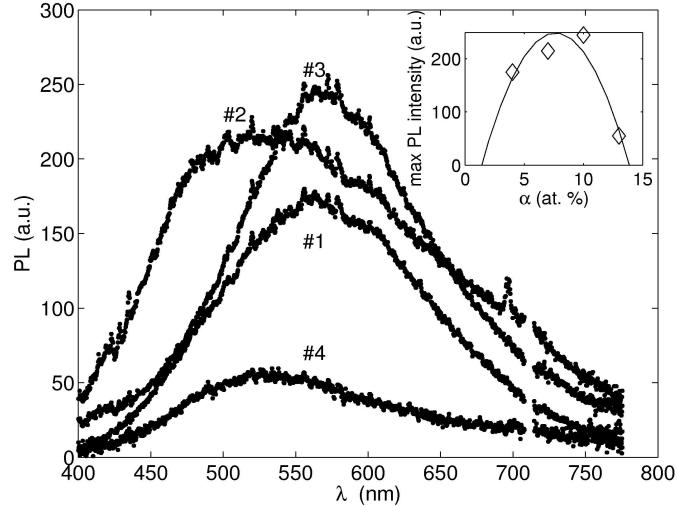

Fig.1. PL spectra of 4 representative samples. #1 -- $Si_{39}O_{36}N_{23}C_2$ ($a$ = 4 at.%); #2 -- $Si_{40}O_{39}N_{18}C_3$ ($a$ = 7 at. %); #3 -- $Si_{44}O_{30}N_{25}C_1$ ($a$ = 10 at. %); #4 -- $Si_{46}O_{26}N_{26}C_2$ ($a$ = 13 at. %). Inset shows PL intensity maximum *vs*. $a$

Fig. 2 shows the effect of the annealing on PL intensity. 30-times growth of PL intensity due to annealing is observed. All annealed samples demonstrate also intensive elecrtoluminescence.

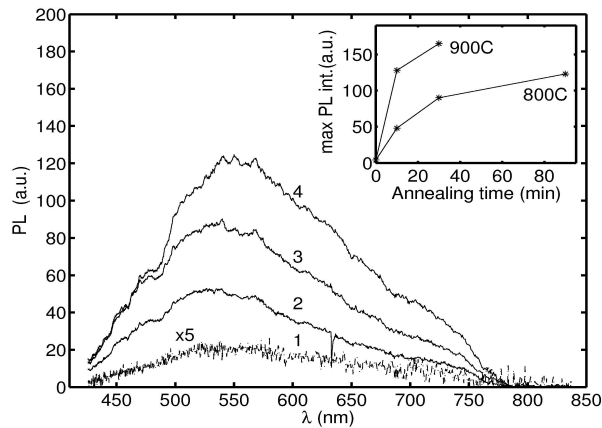

Fig.2. PL spectra at different annealing time (t) of the sample #2: 1 - t = 0, 2 -–10 min, 3 - 30 min, 4 - 90 min (T=800C). Inset shows dependence of maximum PL intensity on annealing time at two annealing temperatures .

Fig. 3 demonstates the strong effect of the annealing on the structure of investigated layers $SiO_xN_y(Si)$. According to TEM data, before annealing the layers where amorphous and slightly non-uniform. Fig. 3 shows high resolution bright-field image of the layer #2 after annealing at 800°C during 30 minutes (cross-section of layer on Si substrate). At the micro-diffraction image the reflects from Si planes (111) with the interplanar distance of 0.312 nm and halo from the layer are presented.

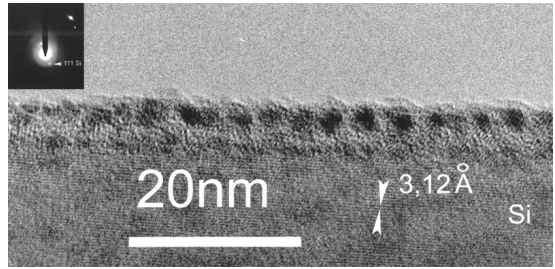

Fig. 3. TEM image of the transverse section of the sample #2 after annealing .

The streak pattern from the atomic planes of Si substrates is seen in the image. On the top of the Si oxynitride layer one can see dark spherical areas (clusters) with sizes of ~3nm and 2-3nm separation in the layer plain.  In some images the streak pattern inside this clasters can be observed, which is an evidence of the atoms ordering.

Fig. 4 shows a typical high resolution spectrum (HRS) of Si atom (level Si2p) obtained by XPS technique for the sample of $SiO_xN_y(Si)$. Analysis of the spectrum allows picking out its components associated with different chemical states of Si atom in oxinitride.

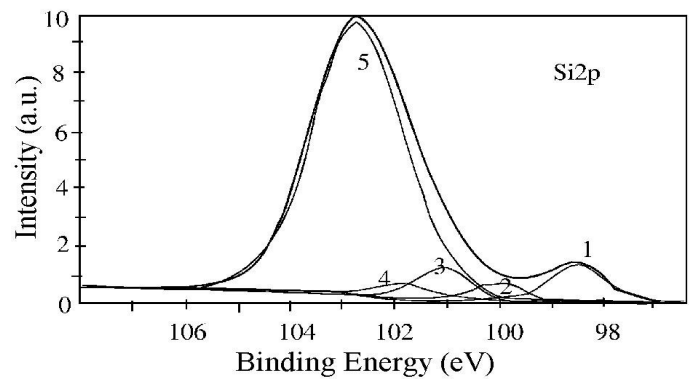

1. Fig. 4. High-resolution XPS spectra of Si2p of annealed sample # 4 ($T = 800^{o}C$, $t = 30$ min).

 The study of these states is important as some of them display luminescence activity.  The basic structural element of stoichiometric Si oxinitride  that defines a short range ordering in this phase is known to be tetrahedron Si-4U with Si atom positioned in the center of tetrahedron and O and N atoms positioned in the 4 vertexes. In oxinitride supersaturated by Si  the excess Si atoms partly replace O and N atoms in the vertexes of some tetrahedrons. Depending of tetrahedron configurations Si atoms turn out to be in different chemical states. These states correspond to the HRS XPS peaks which positions are defined by the binding energies of Si2p level. The binding energy values are obtained from the works [2-4] as well as  from our test measurements.   Fig. 4 shows a decomposition of full HRS of Si atom  for sample #4. Such a decomposition of Si atom HRS has been also made for other samples. Table 1 represents the results of these spectra processing.

Table 1.

| Sample # | Peak 1 $S_{1}/S$, %, $X_{Si}$ at. % | Peak 2 $S_{2}/S$, %, $X_{Si}$ at. % | Peak 3 $S_{3}/S$, %, $X_{Si}$ at. % | Peak 4 $S_{4}/S$, %, $X_{Si}$ at. % | Peak 5 $S_{5}/S$, %, $X_{Si}$ at. % |
|---|---|---|---|---|---|
| 2 | 1,5 | 4,4 | 6,2 | 7,2 | 80,7 |
|   | 0,6 | 1,8 | 2,6 | 3,0 | 32,0 |
| 3 | 2,6 | 4,8 | 7,1 | 5,3 | 80,2 |
|   | 1,2 | 2,1 | 3,2 | 2,2 | 35,3 |
| 4 | 7,4 | 3,7 | 6,1 | 3,1 | 79,7 |
|   | 3,4 | 1,8 | 2,8 | 1,5 | 36,5 |

In the Table, Sn/S means a ratio of the n-th peak square to the full square of a given spectrum, X is a Si content in a corresponding peak. According to the energy peak positions from [4], we can conclude the following:

peak 1 corresponds to a tetrahedron Si-(4Si) silicon phase (Si nanoclusters or nanocrystals),
peak 2 corresponds to a tetrahedron Si-(3Si,N) and/or Si-(3Si,O);
peak 3 corresponds to a tetrahedron Si-(Si,3N) and/or Si-(2Si,O,N);
peak 4 corresponds to a tetrahedron Si-(Si,3O) and/or Si-(Si,2O,N);
peak 5 corresponds to a tetrahedron Si-(2O,2N) and/or Si-(3O,N).
Some of these formations display a luminescence activity; their participation in PL is discussed below.

## 3. Discussion

Thus, a new structure arising after annealing (see Fig.3) is characterized by generation of small Si nanocrystals located in the center of spherical formations observed. Nanocrystal sizes do not exceed 3 nm, the estimation of their surface density gives $3-5 \cdot 10^{12}$ cm$^{-3}$. Previously we have established Si crystallization for similar layers obtained on another substrates (NaCl). It should be noted that according to XPS data (see Fig.4 and Table 1) excess Si after annealing at 800C is concentrated not only in silicon phase (peak 1, Si nanoclusters and nanocrystals) but mainly in chemical complexes in the form of tetrahedrons of 2-4 types (peaks 2-4). These complexes containing excess Si atoms form supersaturated silicon solution in stoichiometric oxinitride (tetrahedrons of type 5). During the subsequent annealing they concentrate around nanoclusters and nanocrystals in interfaces, from where excess Si atoms pass into silicon phase. As follows from Fig.2, the annealing is followed by sharp increase of PL intensity. This behavior is usually associated with the quantum confinement in nanocrystals. However, the main peak in PL spectra observed (see Fig.1 and 2) does not correspond to the energy of radiative transitions in the volume of size-quantized nanocrystals of the same size. According to experimental data, these transitions lie generally in more long-wave region (red or IR) [1,5,12]. In the PL spectra shown on Fig.2 these transitions can be associated with the shoulder in the region $\lambda$=650-750 nm, but their intensity being lower than in the main peak ($\lambda$=500-600 nm). Therefore, the main contribution to the PL observed is likely to be bound with not nanocrystals volume but radiative centers in their boundaries or narrow interfaces. These centers are known (see [6-12]) to be formed on the base of specific structural defects and their complexes associated with excess Si atoms. In the investigated layers these structural defects are represented by tetrahedrons of type 3 Si-(Si,3N) or type 4 Si-(Si,3O) and Si-(Si,2O,N) revealed by XPS technique (see Table 1). In fact, due to tetravalence of Si atom tetrahedrons of types 3 or 4 are formed in couples in the form of $\equiv$Si-Si$\equiv$ complexes, where in every tetrahedron three vertexes are occupied by N or O atoms and the fourth one by Si atom. According to [6-10], these complexes and their associations are luminescent-active; they are responsible for short-range PL ($\lambda$= 400-600 nm) in non-stoichiometric layers $SiO_x$ and $SiO_xN_y$. As follows from Table 1, the small Si excess in investigated layers helps to facilitate the increase of percentage of luminescent-active complexes. During the annealing the number of these complexes and their associations in interfaces can grow and increase the short-range PL intensity. In addition, the transitions between size-quantized energy levels of nanocrystals as well as other groups of defects can make contribution to the long-wave part of PL spectra. It should be also noted that nanocrystals (nanoclusters) can participate actively in an exitation of radiative centers in interfaces. The drastic PL intensity increase at annealing is also related to the improvment of interfaces structure, closing of broken chemical bonds, and effective suppression of non-radiative processes. The increase of Si excess is accompanied with the increase of the number of large closely adjacent Si clusters, the dergadation layer structure, causing a rapid growth of the speed of non-radiative processes and luminescence suppression.

This work was supported by ISTC (project #2556) and by the program of Department of Physical Sciences of Presidium of RAS. We are greateful to S.V. Zaitsev-Zotov for help in experiment.